\documentclass[twocolumn,preprintnumbers,amsmath,amssymb,prl,superscriptaddress]{revtex4}
\usepackage{graphicx}
\usepackage{dcolumn}
\usepackage{bm}
\usepackage{psfrag}
\begin{document}
\renewcommand{\bbox}[1]{{\bm #1}}
\title{Coagulation by Random Velocity Fields as a Kramers Problem}
\author{Bernhard Mehlig}
\affiliation{Physics and Engineering Physics, Gothenburg University/Chalmers, 
Gothenburg, Sweden}
\author{Michael Wilkinson}
\affiliation{Faculty of Mathematics and Computing,
The Open University, Walton Hall, Milton Keynes, MK7 6AA, England}

\begin{abstract}
We analyse the motion of a system of particles suspended
in a fluid which has a random velocity field.
There are coagulating and non-coagulating phases.
We show that the phase transition is related to a Kramers problem,
and use this to determine the phase diagram, as a function of the 
dimensionless inertia of the particles, $\epsilon$, and a measure of 
the relative intensities of potential and solenoidal components of the 
velocity field, $\Gamma$. We find that the phase line is described 
by a function 
which is non-analytic at $\epsilon=0$, and which is related to escape 
over a barrier in the Kramers problem. We discuss the physical
realisations of this phase transition.
\end{abstract}
\maketitle

Deutsch \cite{Deu85} introduced and investigated a model which
can describe the motion of particles suspended in a 
randomly moving fluid. He showed that the 
one-dimensional model exhibits
a phase transition between coagulating and non-coagulating
phases as the effect of inertia of the particles is increased. 
We recently solved Deutsch's one-dimensional model exactly
(\lq path-coalescence model' \cite{Wil03}).

This letter discusses the phase diagram
for the path-coalescence model in higher dimensions,
which is the most relevant case for physical applications. 
In the limit where inertial effects are negligible, the 
suspended particles are advected by the flow: the theory of this
limiting case is described in \cite{Kly99}.
The case where the inertia of the particles play an important role
is much less thoroughly understood. In Ref.~\cite{Bal01} a model is described
in which inertial effects are treated by assuming that
particles are advected in a perturbed velocity field.
Ref. \cite{Bec03} gives numerical results on particle
aggregation in a closely related model.
The aggregation of buoyant particles on the surface 
of a turbulent liquid has also been studied \cite{Sch02}.
Here we give a treatment of the phase transition in higher dimensions, using
an exact mapping to a Kramers problem
(the escape of particles from an attractor by diffusion).

\begin{figure}[tb]
\centerline{\includegraphics[width=6.05cm,clip]{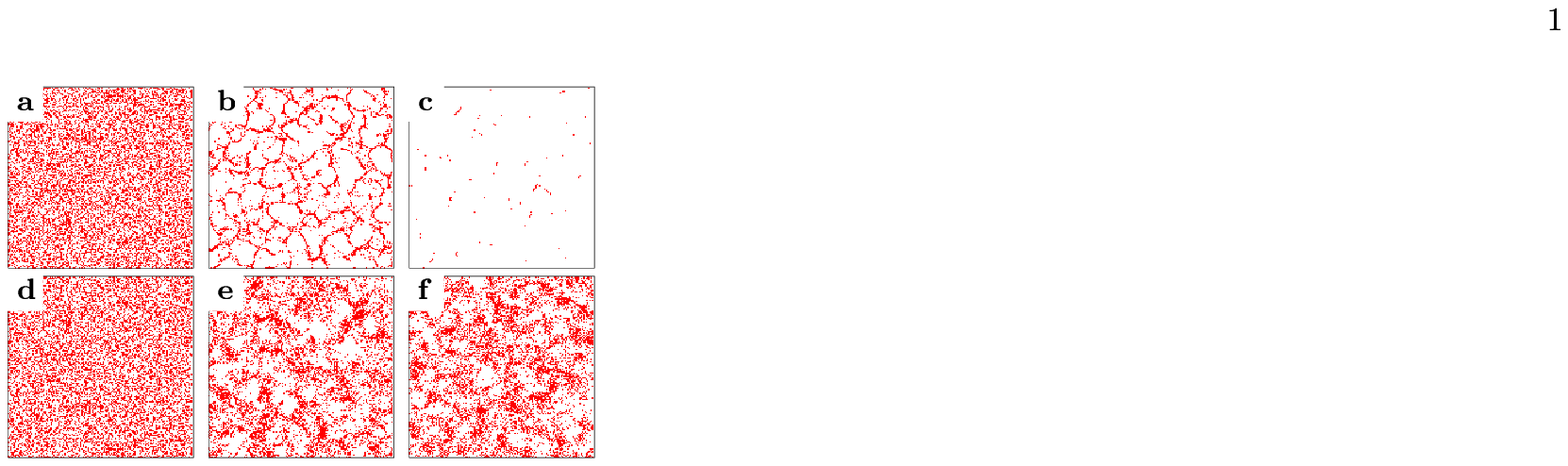}}
\caption{\label{fig: 1} Evolution of uniform initial 
particle distribution under eq. (\ref{eq: 1}) in the unit square
with periodic boundary conditions, 
for $10^4$ particles;  $\Gamma=1/3$, $\epsilon=0.33$ (coalescing phase),
at $t=0$ ({\bf a}), $10$ ({\bf b}),
and $t=275$ ({\bf c}); $\Gamma=1/3$, 
$\epsilon=4.30$ at $t=0$ ({\bf d}), $10$ ({\bf e}), and
$275$ ({\bf f}). The statistics of the force is explained in 
the text, we chose $C(R,\Delta t) 
\propto \exp[-\Delta t^2/(2\tau^2)-R^2/(2\xi^2)]$ with $\xi=0.1/\pi$
and $\tau\rightarrow  0$.}
\end{figure}

In the following we discuss the  phase transition 
in two dimensions (Fig.~\ref{fig: 1}).  
The three-dimensional case is considerably harder
to analyse, but the results are surprisingly similar.
We derive a perturbation expansion for a Liapunov exponent $\lambda_1$
determining the phase transition, in powers
of a dimensionless measure of the inertia, $\epsilon$.
Surprisingly, 
the perturbative result for the phase
line turns out to be 
independent of $\epsilon$. Our numerical simulations,
by contrast, imply that the phase line does depend upon $\epsilon$.
We resolve this apparent inconsistency by showing that 
there is a contribution $\delta\lambda_1$ to $\lambda_1$
which is non-analytic in $\epsilon$, 
characteristic of the flux over a barrier in a Kramers problem: 
$\delta \lambda_1 \sim \exp(-\Phi/\epsilon^2)$, where $\Phi$ is the action
of a trajectory in a Hamiltonian dynamical system.
We conclude with a discussion of physical applications.

We consider non-interacting spherical particles of mass $m$, 
radius $a$, and ${\bbox{r}}(t)$ denotes the position of a typical particle.
These move through a fluid with velocity 
${\bbox{u}}({\bbox{r}},t)$ having viscosity $\eta$.
To avoid complications from displaced-mass effects \cite{Lan59},
we assume
that the particles have much higher density than the surrounding
fluid. Our results are therefore most relevant to suspensions
in gases, and we allow ${\bbox{u}}(\bbox{r},t)$ to be a compressible
flow (by contrast, Refs. \cite{Bal01,Bec03} treat the case
relevant to suspensions in liquids).
We consider the case where the drag force ${\bbox{f}}_{\rm dr}$ on the 
particle is given by Stokes' law: 
${\bbox{f}}_{\rm dr}=6\pi \eta a ({\bbox{u}}-\dot {\bbox{r}})$.
The equation of motion is
\begin{equation}
\label{eq: 1}
\ddot{\bbox{r}}=-\gamma (\dot {\bbox{r}}-{\bbox{u}})
\end{equation}
where $\gamma =6\pi \eta a/m$.
Rearranging (\ref{eq: 1}) gives 
$\dot {\bbox{r}}={\bbox{p}}/m$, 
$\dot {\bbox{p}}=-\gamma {\bbox{p}}+{\bbox{f}}({\bbox{r}},t)$
where ${\bbox{f}}=\gamma m{\bbox{u}}$
will be modelled by a random field.
Linearising to obtain an equation
for the separation $(\delta \bbox{r},\delta \bbox {p})$ 
of two nearby trajectories gives 
\begin{equation}
\label{eq: 3}
\delta \dot {\bbox{r}}=\delta {\bbox{p}}/m
\ ,\ \ \
\delta \dot {\bbox{p}}=-\gamma \delta {\bbox{p}}+{\bf F}(t)\delta
{\bbox{r}}
\end{equation}
where ${\bf F}(t)$ is a $2\times 2$ matrix with elements
$F_{ij}(t)={\partial f_i/{\partial r_j}}({\bbox{r}}(t),t)$.
We write $\delta {\bbox{r}}=X\hat {\bbox{n}}_\theta$ and
$\delta {\bbox{p}}=
Y_1 X\hat {\bbox{n}}_\theta+Y_2 X\hat {\bbox{n}}_{\theta+\pi/2}$,
where $\hat {\bbox{n}}_\theta$ is unit vector in direction $\theta$.
We expect that the scale variable $X$ may increase or decrease, but
that $\theta$, $Y_1$, and $Y_2$ approach
a stationary distribution.

The phase transition is determined
by the behaviour of $X$: if the (maximal) Liapunov exponent 
$\lambda_1=m^{-1}{\rm d}\langle \log_{\rm e}X\rangle/{\rm d}t$ is negative,
the particles coagulate \cite{Wil03}.
Substituting the expressions for $\delta\bbox{r}$ and $\delta\bbox{p}$
into (\ref{eq: 3}), and taking scalar products 
with $\hat {\bbox{n}}_\theta$ and 
$\hat {\bbox{n}}_{\theta+\pi/2}$
we obtain
\begin{equation}
\label{eq: 7}
\dot X=Y_1 X/m
\ ,\ \ \ 
\dot \theta=Y_2/m
\end{equation}
and 
\begin{eqnarray}
\label{eq: 8}
\dot Y_1&=&(Y_2^2-Y_1^2)/m-\gamma Y_1+F_{\rm d}(t)
\nonumber \\
\dot Y_2&=&-2Y_1Y_2/m-\gamma Y_2+F_{\rm o}(t)
\end{eqnarray}
where 
$F_{\rm d}(t)=\hat {\bbox{n}}_{\theta}.{\bf F}(t)\hat {\bbox{n}}_{\theta}$
and 
$F_{\rm o}(t)=\hat {\bbox{n}}_{\theta+\pi/2}.{\bf F}(t)%
\hat {\bbox{n}}_\theta$.
From (\ref{eq: 7}), the distribution of 
$\theta $ becomes uniform on 
$[0,2\pi]$ at large $t$, and the Liapunov exponent is 
\begin{equation}
\label{eq: 9}
\lambda_1=\langle Y_1 \rangle/m
\ .
\end{equation}
The statistics of the random force is assumed to be rotationally
invariant, 
so that the statistics of $F_{\rm d}$ and $F_{\rm o}$ are independent 
of $\theta$.
For $\theta =0$, we have $F_{\rm d}=F_{11}$ and 
$F_{\rm o}=F_{21}$, so we obtain the statistics of $F_{\rm d}$, $F_{\rm o}$ 
from those of $F_{11}$, $F_{21}$.
Eqs.~(\ref{eq: 8}) are thus independent of (\ref{eq: 7}). 
The Liapunov exponent is therefore determined by solving
a pair of coupled stochastic differential equations for $(Y_1,Y_2)$.

To make further progress we must specify the statistical properties
of the isotropic and homogeneous random field ${\bbox{f}}$. 
We consider the case where $\langle {\bbox{f}}\rangle=0$.
Without loss of generality we can write 
\begin{eqnarray}
\label{eq: 10}
{\bbox{f}}({\bbox{r}},t)&=&
\nabla \phi({\bbox{r}},t)+\nabla \wedge \hat{\bbox{n}}_3 \psi ({\bbox{r}},t)
\end{eqnarray}
where $\hat{\bbox{n}}_3$ is a unit vector pointing out of the plane.
The components of ${\bbox{f}}$ arising from $\phi$ and $\psi$ are
termed the potential and solenoidal components, respectively.
In the following 
we assume
that the correlation function $C(R,\Delta t) = \langle
\phi(\bbox{r},t)\phi(\bbox{r}',t')\rangle$ 
is an even function of $\Delta t=t-t'$ with support $\tau$
(the correlation time), and has support $\xi $ 
(the correlation length) in $R=|\bbox{r}-\bbox{r}'|$.
We also assume, for simplicity, that 
the fields $\phi$ and $\psi$ are not correlated with each other,
and have the same correlation functions, apart from
a scale factor $\alpha^2=\langle\psi^2\rangle/\langle\phi^2\rangle$.
These assumptions are easily relaxed.

If the correlation time $\tau $ is short
compared to other relevant time scales
we can write
(\ref{eq: 8}) as a pair of coupled Langevin equations. We scale
these into a dimensionless form, in which the diffusion matrix is the unit
matrix
\begin{eqnarray}
\label{eq: 17}
{\rm d}x_1&=&\big[-x_1+\epsilon(\Gamma x_2^2-x_1^2)\big]{\rm d}t'
+{\rm d}w_1\,,
\nonumber \\
{\rm d}x_2&=&\big[-x_2-2\epsilon x_1x_2\big]{\rm d}t'
+{\rm d}w_2\,.
\end{eqnarray}
Here $x_i = \sqrt{\gamma/D_i}\, Y_i$ (for $i=1,2$), $t'=\gamma t$, 
$\langle {\rm d}w_i\rangle=0$, 
$\langle {\rm d}w_i{\rm d}w_j\rangle=2\delta_{ij}{\rm d}t'$,
and
\begin{equation}
\label{eq: 13}
D_{i}=\frac{1}{2}\!\!\int_{-\infty}^\infty 
{\rm d}t\ \langle F_{i1}(t)F_{i1}(0)\rangle\,.
\end{equation}
The two 
parameters $\Gamma$ and $\epsilon$
are defined as follows:
\begin{equation}
\label{eq: 15}
\Gamma\equiv{D_2/{D_1}}={(1+3\alpha^2)/(3+\alpha^2)}
\end{equation}
characterises the ratio of the solenoidal and potential
field amplitudes,
${1\over 3}\le \Gamma \le 3$, with 
$\Gamma ={1\over 3}$ for purely potential fields, $\Gamma=3$
for pure solenoidal force fields, and $\Gamma =1$ for equal
field intensities. 
The second parameter
\begin{equation}
\label{eq: 16}
\epsilon=D_1^{1/2}/(m\gamma^{3/2})=(mD_1)^{1/2}/(6\pi\eta a)^{3/2}
\end{equation}
is a measure of the importance of inertial effects in the equation of motion.

The Langevin equations (\ref{eq: 17}) are equivalent to a 
Fokker-Planck equation \cite{Ris90}
\begin{equation}
\frac{\partial P}{\partial t'}=\hat{\cal F}P \equiv 
\nabla.[-{\bbox{V}}P+\nabla P] 
\end{equation}
where the advection field ${\bbox{V}}$
has components 
$V_1 = -x_1+\epsilon(\Gamma x_2^2-x_1^2)$ and 
$V_2 = -x_2-2\epsilon x_1x_2$.
We write $\hat{\cal F} = \hat{\cal F}_0 + \epsilon \hat{\cal F}_1$ and
seek a steady-state solution $P$ satisfying
$\hat{\cal F} P = 0$.
The Liapunov exponent $\lambda_1$ is determined by computing 
$\langle x_1\rangle$ with $P(\bbox{x})$, and 
using (\ref{eq: 9}) to obtain 
$\lambda_1=\gamma \epsilon \langle x_1\rangle$.

\begin{figure}[t]
\centerline{\includegraphics[width=6.5cm,clip]{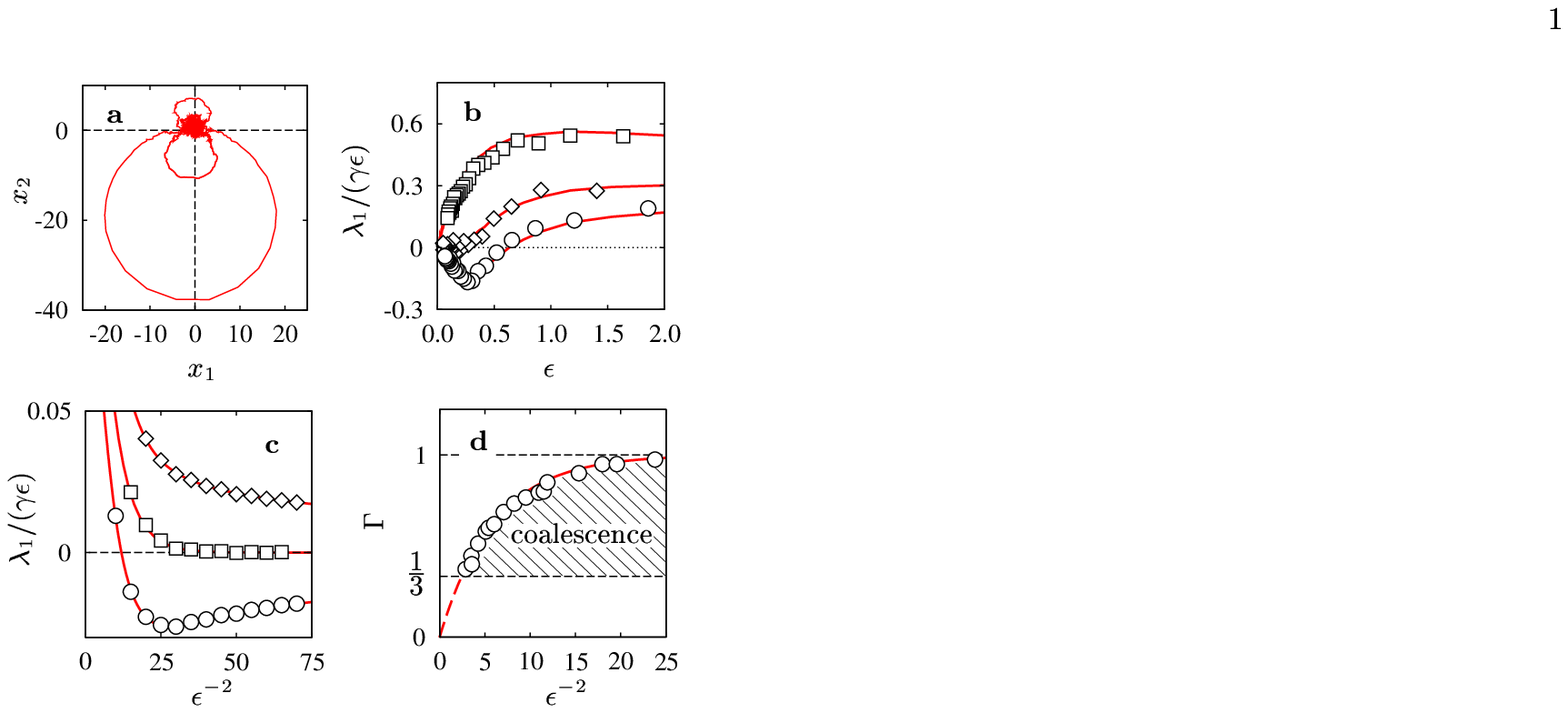}}
\caption{\label{fig: 2}
{\bf a} A trajectory for the Langevin equations, (\ref{eq: 17}), for 
$\epsilon=0.5$, $\Gamma=1$, and $0\leq t'\leq 10^{2}$. 
The trajectory $\bbox{x}(t)$
is most frequently located close to the origin, but occasionally
it passes the unstable fixed point at $(-1/\epsilon,0)$, making a 
large excursion before returning to the origin. The non-analytic
contribution to the Liapunov exponent is associated with these 
large excursions.
{\bf b}
Liapunov exponent as a function of $\epsilon$, for $\Gamma = {1\over 3} (\circ),
1 (\Diamond), 3 (\Box)$: Monte-Carlo simulations of (\ref{eq: 17}) (lines)
and integration of (\ref{eq: 1}) (symbols).
{\bf c}  Comparison of Monte-Carlo simulation of (\ref{eq: 17})
(symbols) and
eq.~(\ref{eq: 33}) (lines), for $\Gamma=0.85 (\circ), 1 (\Box), 1.15 (\Diamond)$
(with $\chi=0.34,0.28,0.215$).
{\bf d} 
Phase diagram in two dimensions: Monte-Carlo simulations ($\circ$)
of (\ref{eq: 17}), the solid line is the function 
$1\!-\!\Gamma=\exp[-1/(6\epsilon^2)]$. 
}
\end{figure}

We start by considering a perturbative approach.
In the limit $\epsilon \to 0$ the steady-state solution is
\begin{equation}
\label{eq: 19}
P_0(\bbox{x})=A\exp[-{\textstyle{1\over 2}}(x_1^2+x_2^2)]
\equiv A\exp[-\Phi_0(\bbox{x})]
\,.
\end{equation}
It is convenient
to make a transformation of the Fokker-Planck operator $\hat{\cal F}$
to a new operator $\hat {\cal H}$ of the form \cite{Ris90}
$\hat{\cal H}=\exp(\Phi_0/2)\hat{\cal F}\exp(-\Phi_0/2)$.
We write 
$\hat{\cal H}=\hat{\cal H}_0+\epsilon \hat{\cal H}_1$ and obtain,
for $\hat{\cal H}_0$,
\begin{equation}
\label{eq: 21}
\hat{\cal H}_0=
(\partial_{x_1}^2-{\textstyle{1\over 4}}x_1^2)
+(\partial_{x_2}^2-{\textstyle{1\over 4}}x_2^2)+1
\end{equation}
where $(\partial_{x_i}^2\!-\!x_i^2/4)$
are one-dimensional quantum-mechanical harmonic-oscillator Hamiltonians 
for coordinates $x_i$. 
The eigenvalues of $\hat{\cal H}_0$ are 
$-(k+l)$, where $k$ and $l$ are non-negative integers. 
We use a variant of Dirac notation denoting the eigenstates of 
$\hat{\cal H}_0$ by $\vert \phi_{kl})$, with 
$(\bbox{x}\vert \phi_{00})=\exp[-\Phi_0(\bbox{x})/2]$,  and
$P(\bbox{x}) = (\bbox{x}|P)$.
Introducing  
annihilation and creation operators, $\hat a_i$ and $\hat a^\dagger_i$,
with the usual properties, 
such that $\hat x_i=\hat a_i+\hat a^\dagger_i$ we find    
\begin{equation} 
\hat {\cal H}_1= \hat a_1^\dagger (\Gamma {\hat x_2}^2-{\hat x_1}^2)
-2\,\hat a_2^\dagger \hat x_2\hat x_1 \,.
\end{equation}
We seek a null eigenvector of $\hat {\cal H}$, 
satisfying $\hat {\cal H}\vert Q)=0$, in the form of a perturbation
series,
$\vert Q)=\vert Q_0)+\epsilon \vert Q_1)
+\epsilon^2\vert Q_2)+...$ where $\vert Q_0)=\vert \phi_{00})$.
The general term satisfies the recursion 
$\hat {\cal H}_0\vert Q_{n+1}) +\hat {\cal H}_1 \vert Q_n)=0$.
In terms of $|Q)$, the expectation of $x_1$ is given by   
\begin{equation}
\langle x_1\rangle =(\phi_{00}|\hat a_1\vert Q )/(\phi_{00}|Q)\,.
\end{equation}
We find that $(\phi_{00}\vert Q)=1$ in all orders of 
perturbation theory. Evaluating $(\phi_{00}\vert \hat a_1\vert Q)$, 
we find that  all of the even orders vanish, and that in the odd orders the 
 square roots arising from the action of annihilation
 and creation operators cancel, so that the perturbation series for
 $\langle x_1\rangle$ has 
 integer coefficients.  We obtain
\begin{equation}
\label{eq: 26}
\lambda_1=-\gamma \sum_{k\geq 1} c_k [(1\!-\!\Gamma)\epsilon^2]^{k}
\end{equation}
with $c_1=1$, $c_2=5$, $c_3=60$, $c_4=1105$, 
$c_5=27120$, $c_6=828250,\ldots$.
This predicts that the Liapunov 
exponent is positive (non-coagulating phase) for $\Gamma >1$ and
negative for $\Gamma <1$, independent of $\epsilon$. 
Fig.~\ref{fig: 2}{\bf b}
shows that this surprising prediction is indeed false.

It is possible that the Liapunov
exponent may have a component non-analytic at $\epsilon=0$,
not captured by perturbation theory.
The series (\ref{eq: 26}) is asymptotic and should
be truncated at the smallest term, with index $k^\ast$.
The coefficients satisfy  the recursion
$c_k=(6k\!-\!8)c_{k-1}+\sum_{l=1}^{k-1}c_lc_{k-l}$ with $c_0=-1/2$, 
and we find
\begin{equation}
\label{eq: 27}
c_k \sim 3^k \sqrt{2k!}
\end{equation}
for $k\rightarrow\infty$,
so that $k^\ast\sim 1+{\rm Int}\{1/[6(1\!-\!\Gamma)\epsilon^2]\}$.
We apply a principle expounded
by Dingle \cite{Din74}, which suggests that the error of an asymptotic series 
is comparable to its smallest term. 
This approach indicates that in the limit 
$\epsilon \to 0$ the non-analytic term is of the 
form $\delta \lambda_1 \sim C\exp\{-1/[6(1\!-\!\Gamma)\epsilon^2]\}$ (where 
$C$ might have a power-law dependence on $\epsilon$). 
This still incorrectly predicts that the non-analytic contribution vanishes at 
$\Gamma=1$, so that the phase
line is independent of $\epsilon$.

We have therefore used an alternative approach:
we write $P$ as
$P=A\exp(-\Phi)$ so that $\Phi$ must satisfy
\begin{equation}
\label{eq: 28}
\nabla^2 \Phi -\nabla.{\bbox{V}}-{\bbox{V}}.\nabla \Phi-(\nabla \Phi)^2=0
\ .
\end{equation}
The deterministic advection 
velocity field ${\bbox{V}}$ contains terms which are quadratic in 
${\bbox{x}}$.
This suggests that $\Phi$ is bounded by a cubic in ${\bbox{x}}$.
In (\ref{eq: 28}), the first two terms are 
expected to be bounded by multiples of
$\vert {\bbox{x}}\vert$, whereas the remaining terms are expected
to be bounded by a cubic function of $|{\bbox{x}}|$. We expect 
that close to the origin, the solution is well approximated by
$\Phi\sim \Phi_0={\textstyle{1\over 2}}(x_1^2+x_2^2)$, and
that far from the origin $\Phi$ is 
well approximated by the solution of the equation containing
only the leading-order terms, i.e.
${\bbox{V}}.\nabla \Phi +(\nabla \Phi)^2=0$.
This has the form of a Hamilton-Jacobi equation 
$H({\bbox{x}},\nabla \Phi)=E$, where in our case $E=0$ and where the 
Hamiltonian is \cite{Fre84}
$H({\bbox{x}},{\bbox{p}})={\bbox{V}}({\bbox{x}}).{\bbox{p}}+{\bbox{p}}^2$.
The Hamilton-Jacobi equation is solved 
by integrating Hamilton's equations, then $\Phi$ is obtained 
by integration along the trajectories:
$\Phi({\bbox{x}})=\int^t {\rm d}t\ \dot{\bbox{x}}.{\bbox{p}}$.
We start classical trajectories infinitesimally close to the origin
with initial conditions ${\bbox{p}}=\nabla \Phi={\bbox{x}}$,
integrate them numerically,
investigate the form of the 
trajectories, and determine the action function
$\Phi(\bbox{x})$. We are especially interested in singularities
of $\Phi$, because these are expected to lead to non-analytic
behaviour of the Liapunov exponent.

On the $x_1$-axis, for $p_2=0$, the vectors
$\dot {\bbox{x}}$, ${\bbox{V}}$ and 
${\bbox{x}}$ are parallel, and ${\bbox{p}}=-{\bbox{V}}$. 
This trajectory crosses a singularity at the point 
${\bbox{x}}^\ast=(-1/\epsilon,0)$, at which both
${\bbox{V}}$ and ${\bbox{p}}$ vanish, with action 
$\Phi(-1/\epsilon,0)=1/(6\epsilon^2)$.
For $0<\Gamma <2$ our numerical experiments 
did not identify any other singular point
with a smaller value of $\Phi$. Consider the physical significance
of this singular point in terms of the dynamics of a fictitious particle
with position $\bbox{x}(t)$ described by the 
Langevin equation (\ref{eq: 17}). A
typical trajectory is plotted in Fig.~\ref{fig: 2}{\bf a}. 
The advection field $\bbox{V}$
has an unstable fixed point at ${\bbox{x}}^\ast$. 
To the right of the singularity,
the particle is advected back towards the origin, and the 
probability density decreases rapidly as the fixed point is approached
from that direction.
To the left of the singularity, the particle is advected away, following 
the advection field $\bbox{V}$: initially it moves to the left, 
returning to large positive $x_1$ in a large circuit.
The singularity is therefore
associated with diffusive escape from the attractor of ${\bbox{V}}$
at ${\bbox{x}}={\bbox{0}}$, in which the 
escaping particle is initially advected away, but 
returns along paths close to the positive
$x_1$ axis. This leads to the hypothesis that there should be a 
contribution to $\lambda_1$ proportional to
$\exp[-\Phi({\bbox{x}}^\ast)]= \exp[-1/(6\epsilon^2)]$,
where the pre\-factor may have an algebraic dependence on 
$\epsilon$.

According to our numerical results, the dominant correction
to the perturbation series in the limit $\epsilon \to 0$ 
does indeed arise in this way: we find
\begin{equation}
\label{eq: 33}
{\lambda_1/{\gamma}}\sim \chi(\Gamma)\exp[-1/(6\epsilon^2)]
-\sum_{k=1}^{k^\ast}
c_k[(1\!-\!\Gamma)\epsilon^2]^{k}\,.
\end{equation}
We have not been able to determine the form of the prefactor
$\chi$ analytically: it appears to be independent of $\epsilon$.
In Fig.~\ref{fig: 2}{\bf c}
we compare the Monte-Carlo simulations of (\ref{eq: 17})  with 
(\ref{eq: 33}). 
Fig.~\ref{fig: 2}{\bf d}  shows the phase diagram, as determined using
(\ref{eq: 9}) and Monte-Carlo simulation of (\ref{eq: 17}).
The computed curve is compared with an empirical fit, using the 
exponential function $\exp[-1/(6\epsilon^2)]$ which
characterises the escape process.

The coefficients in (\ref{eq: 26}) also occur in the asymptotic
expansion of ${\rm Ai}'(z)/{\rm Ai}(z)$ \cite{Pra01}. This suggests that
${\lambda_1/{\gamma}}=
-\mbox{Re}\,[{{\rm Ai}'(z)/{{\rm Ai}(z)}}+\sqrt{z}]/(2\sqrt{z})$
with $z= ({\rm i}\sqrt{1\!-\!\Gamma}\epsilon)^{-4/3}/4$.
This expression cannot be correct for $\vert 1\!-\!\Gamma\vert<1$,
because its leading order non-analytic correction becomes smaller
than the non-analytic term in (\ref{eq: 33}) in the limit 
$\epsilon \to 0$. However, on setting $\Gamma=0$,we find
that it does agree with the exact solution of the one-dimensional
problem obtained in \cite{Wil03}, and our numerical results show
excellent agreement with Monte-Carlo simulations when $\Gamma=2$
(where we believe that it could also be exact), and when $\Gamma=3$.

In summary, we have seen that the theory of coagulation by random velocity
fields bears several surprises. 
We have seen that the phase transition is determined by the 
stationary state of a Langevin process. 
Perturbation theory incorrectly predicts
that the phase line is independent of the inertia parameter
$\epsilon$. The asymptotics of the high order terms of the 
perturbation series again do not predict the correct form of the
non-analytic term; we are not aware of any other physical
example where this procedure fails (although a mathematical
example was suggested in \cite{Bal79}). 
Finally, the path-coalescence phase transition
is driven by a non-analytic term
characterising the flux of a barrier in a Kramers problem.

\begin{figure}[b]
\centerline{\includegraphics[width=6.05cm,clip]{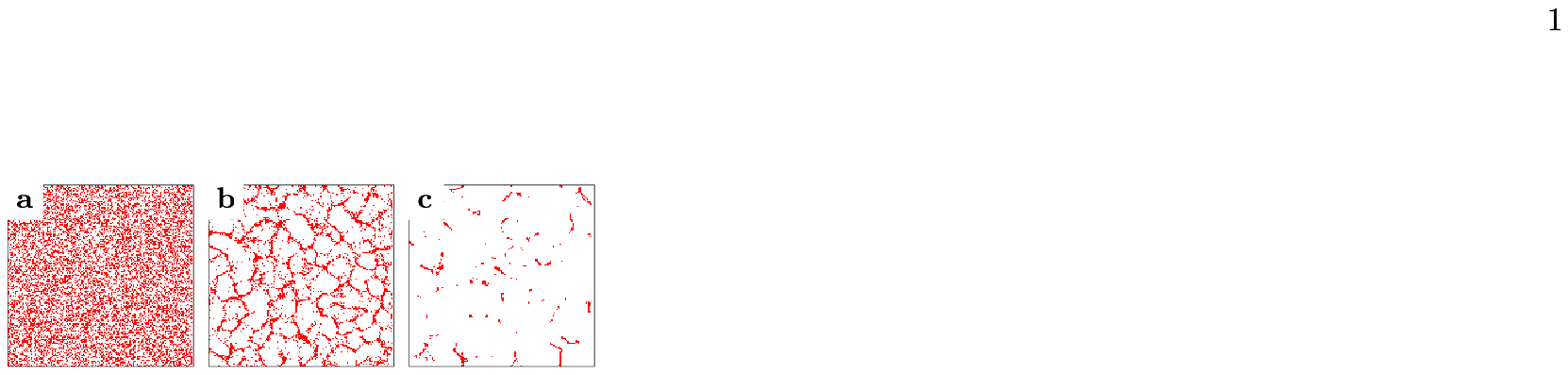}}
\caption{\label{fig: 3} Same as Fig.~\ref{fig: 1}{\bf a} - {\bf c}, but 
with masses uniformly distributed on an interval of half width 
$\delta m=0.2\langle m\rangle$.
}
\end{figure}

We conclude with a number of remarks and a discussion of
possible applications of the effect.
First, in many applications the 
suspended particles will not have equal masses, and it is necessary to
consider how mass dispersion affects the coagulation process.
We have shown, in a perturbative framework, that two particles 
with masses differing by $\delta m\ll m$
follow trajectories with separation 
$\Delta {\bbox{r}}=\delta m\, {\bbox{g}}(t)$, where 
$\langle \vert {\bbox{g}}(t)\vert^2\rangle$ remains bounded
as $t\to \infty$. We infer that when the masses are
unequal, particles condense onto fragmented line segments
rather than isolated points. 
The coagulation effect is therefore weakened, but
not destroyed (Fig.~\ref{fig: 3}) by mass dispersion.
Second, the structures observed in Fig.~\ref{fig: 1}{\bf b} 
indicate that the area-contraction rate is much larger
than the maximal Liapunov exponent $\lambda_1$.
We have verified this by computing
the Liapunov spectrum 
$\lambda_4 \!<\! \lambda_3\! <\! \lambda_2\! <\! \lambda_1$ of (\ref{eq: 1}) 
numerically
(the area-contraction rate is given by $\lambda_2+\lambda_1$).
Similar structures were observed in computer simulations
of inertial particles in a chaotic flow \cite{Bec03}. 
In two dimensions, $\lambda_2$ can be
found from the stationary state of Langevin equations similar in
structure to (\ref{eq: 17}).

Turning to specific applications, we note that the random velocity
field ${\bbox{u}}({\bbox{r}},t)$ 
could arise from random sound waves, turbulent
flow, or other random disturbances. We do not
know how to estimate $\Gamma$ reliably for turbulent flow in
gases. In the case of liquids where the flow is incompressible,
our theory applies when the suspended particles are denser
that the fluid: we predict there is no coagulation because $\Gamma >1$.
In sound waves the velocity field is proportional to the pressure 
gradient, and is therefore pure potential: $\Gamma={1\over 3}$, 
so according to Fig.~\ref{fig: 2}{\bf d} there is a coagulating phase. 
One possible technological application of the effect would be to the
coagulation of small pollutant particles in an engine exhaust. An 
ultrasonic noise source could increase the
size of the pollutant particles until they are large enough to
be captured efficiently by a mechanical filter.
Coagulation by ultrasonic sources has been observed experimentally
\cite{Goo69}. Theoretical treatments, reviewed in \cite{Goo69},
have considered the case where the ultrasound has a single frequency,
and the coagulation results from particles with differing masses
experiencing different displacements. This letter has introduced a new
mechanism for ultrasonic coagulation, which works even when particles
have the same mass.


\begin{thebibliography}{6}


\bibitem{Deu85}
J. M. Deutsch, {\sl J. Phys. A: Math. Gen.} {\bf 18} 1457 (1985).


\bibitem{Wil03}
M. Wilkinson and B. Mehlig, {\sl Phys. Rev. E} {\bf 68}  040101(R) (2003).


\bibitem{Kly99}
V. I. Klyatskin and D. Gurarie, {\sl Physics Uspekhi} {\bf 42} 165 (1999).


\bibitem{Bal01}
E. Balkovsky, G. Falkovitch, and A. Fouxon, {\sl Phys. Rev. Lett.} {\bf 86}
2790 (2001).

\bibitem{Bec03}
J. Bec, nlin/0306049 (2003).

\bibitem{Sch02}
J\"org Schumacher and B. Eckhardt, Phys. Rev. E {\bf 66} 017303 (2002).

\bibitem{Lan59}
L. D. Landau and E. M. Lifshitz, {\sl Fluid Mechanics}, Pergamon: Oxford (1959).

\bibitem{Ris90}
H. Risken, {\sl The Fokker-Planck Equation. Methods of Solutions and
Applications}, Springer: New York (1999).

\bibitem{Din74}
R. B. Dingle, {\sl Asymptotic Expansions: their Derivation and Interpretation}, 
Academic: New York (1974).

\bibitem{Fre84} M. I. Freidlin and A. D. Wentzell, {\sl Random Perturbations
of Dynamical Systems}, Springer: New York (1984).

\bibitem{Pra01}
M. Pr\" ahofer, in: {\sl On-Line} {\sl Encyclopaedia} {\sl of} {\sl Integer} {\sl Sequences}, A062980 (2001).

\bibitem{Bal79} R. Balian, G. Parisi, and A. Voros, in: Lecture Notes in
Physics {\bf 106}, Springer: New York (1979).

\bibitem{Goo69} G. L. Gooberman, {\sl Ultrasonics: Theory and Application}, 
Hart Publications: New York (1969).

\end{thebibliography}
\end{document}